\documentclass[11pt]{article}

\usepackage[
  top=2.5cm,
  bottom=2.5cm,
  left=2.5cm,
  right=2.5cm
]{geometry}

\usepackage[utf8]{inputenc}
\usepackage[T1]{fontenc}
\usepackage{amsmath,amssymb,amsfonts}
\usepackage{mathtools}
\usepackage{graphicx}
\usepackage{booktabs}
\usepackage{xcolor}
\usepackage{hyperref}
\usepackage{cleveref}
\usepackage{float}
\usepackage{caption}
\usepackage{subcaption}
\usepackage{enumitem}
\usepackage{microtype}
\usepackage{setspace}

\usepackage[
  backend=biber,
  style=nature,
  sorting=none,
  maxbibnames=10,
  minbibnames=3
]{biblatex}
\usepackage{pifont}
\newcommand{\cmark}{\textcolor{green!60!black}{\ding{51}}}
\newcommand{\xmark}{\textcolor{red!70!black}{\ding{55}}}
\newcommand{\pmark}{\textcolor{yellow!50!orange!90!black}{\Large\textbullet}}

\hypersetup{
  colorlinks=true,
  linkcolor=blue!60!black,
  citecolor=blue!60!black,
  urlcolor=blue!60!black
}

\newcommand{\modelname}{HiPoly}
\newcommand{\grins}{G\textsuperscript{2}RINS}%{GGRINS}

\begin{document}
% ============================================================================

% --- Title ---
\begin{center}
  {\LARGE\bfseries
    HiPoly: a hierarchical polymer-native AI framework for\\
    property prediction and generative design
  }
  \vspace{1em}

  % --- Authors ---
  {\large
    Ge Sun$^{1,2}$,
    Gervasio Zaldivar$^{1,\dagger}$,
    Yuan Tian$^{1,\dagger}$,
    Gustavo Perez Lemus$^{1}$,
    Juhae Park$^{1,4}$,
    Dasha Safarian$^{4}$,
    Ming Han$^{5}$,
    Juan J. de Pablo$^{1,2,3,*}$
  }
  \vspace{0.5em}

  % --- Affiliations ---
  {\small
    $^{1}$Department of Chemical and Biomolecular Engineering, Tandon School of Engineering, New York University, Brooklyn, NY, USA\\
    $^{2}$Department of Computer Science, Courant Institute of Mathematical Sciences, New York University, New York, NY, USA\\
    $^{3}$Department of Physics, New York University, New York, NY, USA\\
    $^{4}$Pritzker School of Molecular Engineering, University of Chicago, Chicago, IL, USA\\
    $^{5}$Center for Quantitative Biology and Peking-Tsinghua Joint Center for Life Sciences, Academy for Advanced Interdisciplinary Studies, Peking University, Beijing, China\\[0.5em]
    $^{*}$Corresponding author: jjd8110@nyu.edu
  }
\end{center}

\let\thefootnote\relax
\footnotetext{$^\dagger$Equal contribution.}

\vspace{1em}

% ---------------------------------------------------------------------------
% Abstract
% ---------------------------------------------------------------------------

\begin{abstract}
Polymeric materials are central to modern technologies, 
with applications ranging from 
energy to health and transportation. 
Although AI has made significant advances in materials discovery, 
the hierarchical structure of polymers across multiple 
length scales makes them inherently difficult to 
represent in a unified and physically meaningful way.
Here we introduce \modelname{}, a
polymer-native AI framework that processes complete polymer
descriptions through a three-level hierarchical graph architecture
built on the \grins{} representation.
\modelname{} encodes stochastic inter-monomer connectivity, composition,
and molecular weight directly within its architecture, using
physically motivated design principles that mirror the multi-scale
nature of polymeric systems.
The framework establishes an end-to-end AI-driven workflow from
experimental formulation data to property prediction, generative
molecular design, and physics-based validation through molecular
simulations, all unified by a single polymer representation.
We demonstrate state-of-the-art prediction accuracy for
thermophysical properties of multi-component polymer systems, with
ablation studies confirming that each hierarchical design choice
contributes independently to model performance.
As an example, the generative design pathway is applied here to the discovery of
sustainable alternatives to persistent fluorinated polymers, where it is possible to 
identify and independently validate PFAS-free candidates with
target surface-energy properties. This work demonstrates how polymer-native AI
can accelerate discovery by linking representation, 
prediction, and design across complex polymer chemistries.
\end{abstract}

\vspace{1em}

% ============================================================================
% MAIN TEXT
% ============================================================================

% ---------------------------------------------------------------------------
% INTRODUCTION
% ---------------------------------------------------------------------------

\section*{Introduction}
\label{sec:introduction}

\begin{figure}[!ht]
  \centering
  \includegraphics[width=\textwidth]{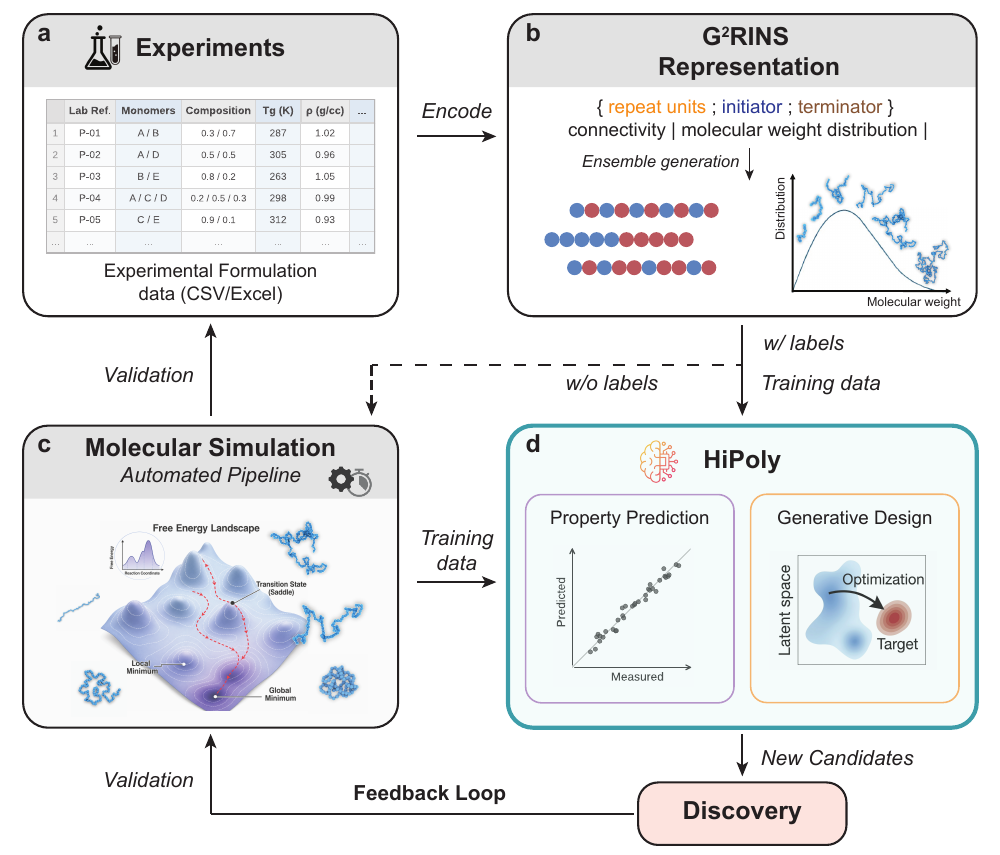}
  \caption{%
    \textbf{End-to-end workflow for polymer property prediction and
    molecular design.}
    \textbf{a},~Experimental formulation data, including monomer
    identities, compositions, and measured properties, serve as the
    starting point for the workflow.
    \textbf{b},~Formulations are encoded into \grins{} polymer strings
    that capture monomer chemistry, mole fractions, stochastic
    inter-monomer bond connectivity, and molecular weight distribution;
    chain ensembles are generated directly from the \grins{}
    specification.
    \textbf{c},~An automated molecular dynamics (MD) simulation pipeline constructs,
    equilibrates, and extracts properties from \grins{}-generated
    chain ensembles, providing training labels when experimental
    measurements are unavailable and independently validating
    generative candidates.
    \textbf{d},~\modelname{} accepts the hierarchical molecular graphs
    derived from \grins{} and performs multi-target property prediction
    or generative molecular design through a shared latent space.
    New candidates identified through generative design feed back into
    simulation or experiment for validation, closing a bidirectional
    feedback loop that progressively augments the training set.
  }
  \label{fig:workflow}
\end{figure}

Polymers are used in applications
spanning energy storage, structural engineering, biomedicine, and
environmental remediation, to name a few~\cite{tran2024review,matyjaszewski2009}.
However, their complexity sets them apart from other material
classes: polymer properties emerge not from a single molecular
structure but from an ensemble of chains whose behavior is governed
by chemistry on multiple scales~\cite{flory1953,rubinstein2003polymer}, from monomer identity and functional
group arrangement at the atomic level to molecular weight, chain
architecture, and monomer sequence statistics at the chain level.
This point is illustrated in Fig.~\ref{fig:workflow}b, which depicts
an ensemble of chains for a polymer. As can be seen in the figure, 
a macroscopic sample is not a single chain, 
but rather a collection of order $10^{23}$ structurally
distinct molecules.
These hierarchical dependencies cannot be captured by monomer
structure alone~\cite{bicerano2002,odian2004}.
The challenge is further compounded in multi-component systems, which
constitute the majority of industrially relevant formulations~\cite{bates2012multiblock}, where
properties depend jointly on composition~\cite{bates1999block,badi2013microstructure}, stochastic inter-monomer
connectivity~\cite{palermo2012impact,lutz2013sequence}, and molecular weight distribution~\cite{gentekos2019controlling,walsh2020general}, creating a
combinatorial design space that far exceeds what experiment or
simulation can explore~\cite{de2019new}.

Artificial intelligence (AI) and machine learning (ML) have emerged
as powerful tools to navigate this space, and the field of
polymer informatics has advanced significantly in recent
years~\cite{audus2017,polymer_genome2018,batra2021review,tran2024review,ge2025review}.
However, most of the existing approaches operate on monomer-level
representations.
Fingerprint-based methods encode individual repeat units as
fixed-length descriptors, handling multi-component systems through
composition-weighted sums or auxiliary
features~\cite{rogers2010ecfp,kuenneth2021copolymer,shukla2024beyond_homopolymers}.
Pre-trained language models such as
polyBERT~\cite{kuenneth2023polybert},
TransPolymer~\cite{transpolymer2023}, and
polyBART~\cite{polybart2025} learn expressive representations from
polymer SMILES strings and can achieve good performance on
homopolymer benchmarks, but their inputs remain monomer-level
sequences that do not natively encode composition or inter-monomer
connectivity.
Graph neural network approaches have moved closer to
polymer-level representations:
Polymer Chemprop~\cite{aldeghi2022wdmpnn} introduced weighted edges
to capture monomer stoichiometry and architecture type in a flat
molecular graph;
PolymerGNN~\cite{polymergnn2023},
CoPolyGNN~\cite{copolygnn2025}, and related multitask GNN frameworks~\cite{gurnani2023multitask} aggregate monomer-level graph
embeddings with composition-aware readout functions;
and self-supervised strategies have been explored to mitigate data
scarcity in polymer GNNs~\cite{gao2024selfsupervised}.
Topology-aware and physics-guided models have been developed to
address architectural diversity in polymer systems, but these operate 
at coarse-grained resolution and do not capture
key atomistic chemical details~\cite{jiang2024topology_vae,jiang2025physics_guided}.
Recent work has shown that decomposing molecules into functional group motifs, 
rather than treating them as flat atom graphs, 
captures chemically meaningful substructures, leading 
to considerable improvements in property prediction~\cite{han2025}. 
Despite these advances, no existing model natively encodes the full
specification of a multi-component polymer, including monomer
chemistry, mole fractions, stochastic inter-monomer bond
connectivity, and molecular weight distribution, within a single
architecture.

On the generative front, the inverse design of new polymers is still in its infancy:
existing approaches target molecular topology at coarse-grained
resolution~\cite{jiang2024topology_vae}, polymer backbones without
composition control~\cite{polyg2g_2021}, or operate on
string-based representations that limit structural
diversity~\cite{polymer_gen_rl_2025,vogel2025inverse}.
None couples generation with physics-based validation through a
shared polymer representation.
 
A key enabler of the present work is the 
Generative Graph Representation of Integrated
Nested BigSMILES (\grins{})~\cite{grins}, 
a polymer representation that builds on the
BigSMILES line notation for stochastic
macromolecules~\cite{bigsmiles2019} and its generative
extension, G-BigSMILES~\cite{gbigsmiles2024}. 
\grins{} captures monomer chemistry, mole
fractions of monomers within polymer chains (or composition), stochastic inter-monomer bond connectivity, and molecular
weight distribution in a single string.
Because \grins{} specifies the complete polymeric system, it can
natively generate atomistic chain ensembles
suitable for molecular dynamics (MD) simulation, providing a unified
interface between ML and physics-based modeling.

Here we introduce \modelname{}, a Hierarchical Generative Polymer AI framework
that operates on
true polymer representations through a multi-resolution architecture
built on \grins{}.
\modelname{} processes the polymer structure at three levels of resolution, from
atoms through functional group motifs to monomer-level connectivity,
using physically motivated design principles: stochastic edge
weighting derived from actual polymerization statistics,
composition-aware aggregation, hierarchical context integration
through root vectors, and molecular weight transformation.
The same architecture supports both multi-target property prediction
and variational generative design, with candidates validated through
MD simulations using \grins{}-generated chain ensembles.
We demonstrate this framework in the prediction of the glass transition
temperature and density for multi-component polymer systems, where
\modelname{} achieves state-of-the-art accuracy from a modestly
sized training set, and in the generative design of PFAS-free
polymer alternatives with MD-validated surface-energy performance.

% ============================================================================
% Results Section (Brief Communication / Letter)
% ============================================================================
% 3 main figures + 2 Tables:
%   Fig 1: Overview (workflow + example + dataset)
%   Fig 2: Architecture (three-level encoder + design principles)
%   Fig 3: Generative design + MD validation (merged)
%   Tab 1-2: Prediction performance + baseline comparison
% Ablation study: Extended Data (referenced from Methods)
% Uncertainty quantification: Extended Data

% ---------------------------------------------------------------------------
% Figure 1: End-to-end workflow
% ---------------------------------------------------------------------------

% ---------------------------------------------------------------------------
\section*{Results}
% ---------------------------------------------------------------------------
\subsection*{An end-to-end AI framework for polymer discovery}
\label{sec:results-overview}
% ---------------------------------------------------------------------------

Predicting the properties of polymeric materials from their molecular
formulation remains a central challenge in materials science,
particularly for hetero-polymer systems where monomer identity,
composition, inter-monomer connectivity, and molecular weight jointly
govern the macroscopic behaviour.

\modelname{} addresses this challenge through a unified workflow that
connects experimental formulation data to property prediction,
generative molecular design, and physics-based validation
(Fig.~\ref{fig:workflow}).

The workflow begins by encoding experimental or simulation records of
monomer identities, compositions, and measured properties
(Fig.~\ref{fig:workflow}a) into \grins{} polymer
strings~\cite{grins}, a polymer-native representation that captures
monomer chemistry, mole fractions, stochastic inter-monomer bond
connectivity, and molecular weight distribution
(Fig.~\ref{fig:workflow}b).
This encoding translates polymer formulation data from human-readable
tabular formats into a machine-readable, graph-constructible form
that preserves the complete structural and statistical specification
of the polymeric system.
A full description of \grins{} is provided in ref.~\cite{grins}.
The resulting \grins{} strings are then parsed into hierarchical
molecular graphs through a functional group
decomposition~\cite{han2025} that coarse-grains each monomer
fragment from individual atoms up to chemically meaningful motifs
such as ring systems, ether linkages, and ester groups, while
inter-monomer connections are annotated with \grins{}-derived bond
probabilities that reflect the stochastic nature of polymerization.

From these hierarchical graphs the framework operates in three stages
(Fig.~\ref{fig:workflow}c,d).
First, a hierarchical graph neural network encoder produces a
polymer-level embedding that jointly captures atom-level chemistry,
functional group organization, and monomer composition
(Fig.~\ref{fig:workflow}d); the architecture of this encoder is
detailed in the next section and Fig.~\ref{fig:architecture}.
Second, this embedding feeds either a multi-target regression head
for property prediction or a shared variational latent space for
generative molecular design.
Third, \grins{}-generated chain ensembles provide the starting
configurations for MD simulations (Fig.~\ref{fig:workflow}c), which
serve a dual role: they supply training labels when experimental
measurements are unavailable, and they independently validate
candidates produced by the generative model.
This closes a bidirectional feedback loop in which new candidates
identified through generative design are confirmed by simulation or
experiment, and the resulting data in turn augment the training set
for subsequent model iterations (Fig.~\ref{fig:workflow}, bottom).
Because \grins{} natively generates atomistic chain ensembles
suitable for simulation, the same representation serves both as
input to the neural network and as the starting point for MD,
ensuring self-consistency across the entire pipeline.

We applied this framework to a dataset of 65
multi-component polymer systems spanning five
polymer families, with up to four distinct monomer
types per formulation.
The target properties include the glass transition temperature $T_g$ and
the mass density $\rho$, covering a wide range of linear and branched
architectures at varying composition fractions.
Rather than compiling property values from literature reports that
may reflect different measurement conditions, we generated all
target properties through MD simulations using \grins{}-generated
chain ensembles and an automated simulation
pipeline for system construction, equilibration,
and property extraction (Fig.~\ref{fig:workflow}c).
This strategy ensures that all properties in the dataset are computed
under identical thermodynamic conditions and simulation protocols,
eliminating a major source of noise in polymer property databases and
providing a self-consistent testbed for model development.
Data were split into training, validation, and test sets; details are
provided in Methods.

% ---------------------------------------------------------------------------
\subsection*{Physically motivated architecture}
\label{sec:results-architecture}
% ---------------------------------------------------------------------------

\begin{figure}[!ht]
  \centering
  \includegraphics[width=\textwidth]{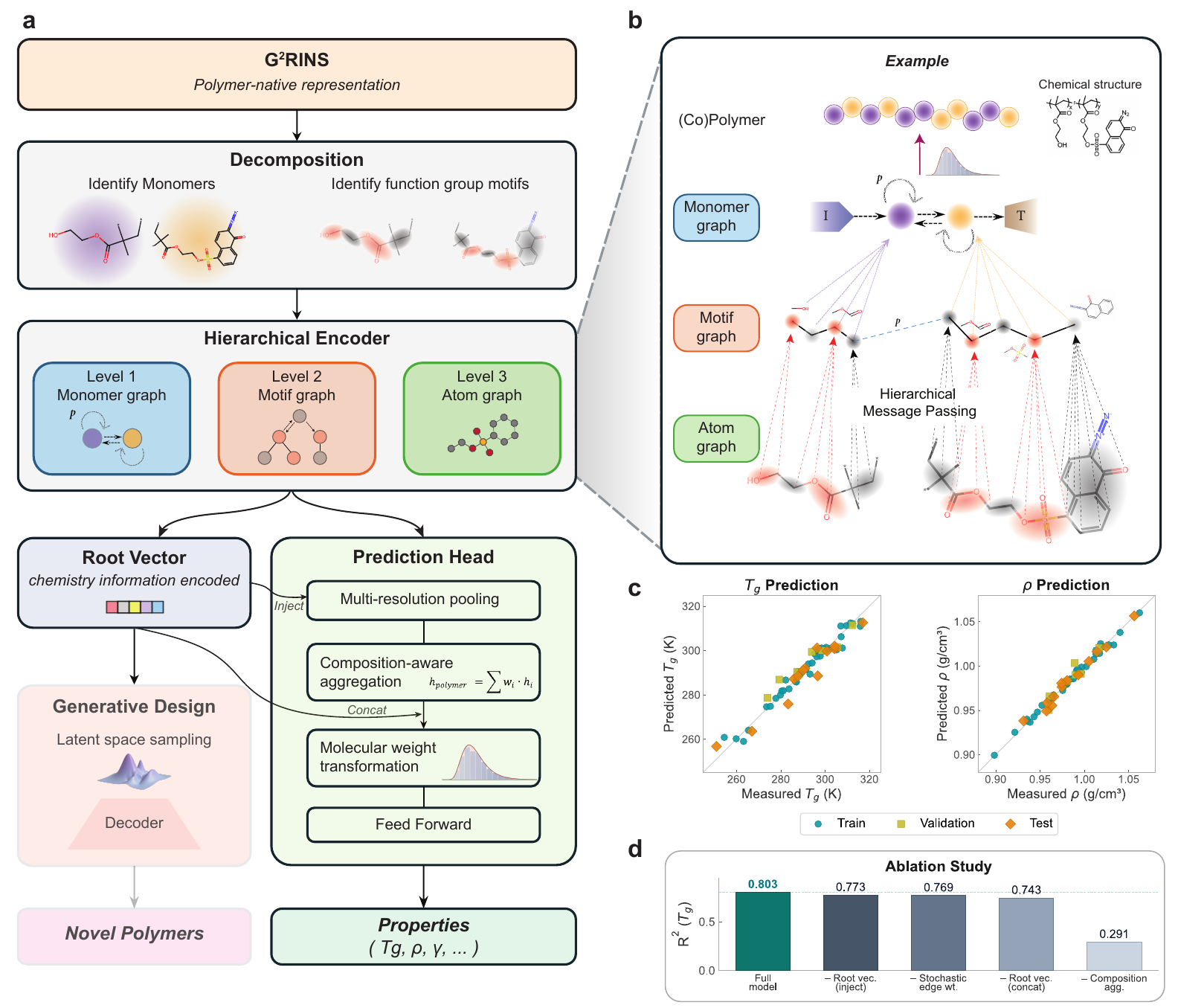}
  \caption{%
    \textbf{\modelname{} architecture, prediction performance, and
    ablation study.}
    \textbf{a},~Overview of the \modelname{} architecture.
    A \grins{} string is decomposed into three graph levels
    (monomer, motif, and atom), each processed by directed
    message passing.
    A root vector summarizes global chemistry information and connects
    to both the prediction head and a variational latent space for
    generative design.
    \textbf{b},~Example of hierarchical message passing on a
    multi-component polymer with initiator (I) and terminator (T)
    fragments, illustrating how information flows across the atom,
    motif, and monomer graph levels.
    \textbf{c},~Parity plots for $T_g$ and density predictions 
    from one representative cross-validation fold; 
    full five-fold statistics are reported in Table~\ref{tab:baselines}.
    \textbf{d},~Ablation study on $T_g$ $R^2$.
    Each variant removes exactly one component from the full model. 
    All components independently contribute to prediction accuracy.
  }
  \label{fig:architecture}
\end{figure}

The \modelname{} architecture encodes the polymer structure in three
resolutions that mirror the physical hierarchy of polymeric materials
(Fig.~\ref{fig:architecture}a).
A \grins{} string is first decomposed by identifying its constituent
monomer fragments and their functional group
motifs~\cite{han2025}, which yields three graph levels.
The atom graph (Level~3, finest resolution) represents each monomer
fragment as a molecular graph of covalent bonds.
The motif graph (Level~2) captures relationships between chemically
meaningful functional groups, such as ring systems and ester groups, within each fragment.
The monomer graph (Level~1, coarsest resolution) encodes the
connectivity among monomer fragments, with edges annotated by
\grins{}-derived bond probabilities that reflect the stochastic nature
of polymerization.
All three levels are processed by directed message
passing~\cite{yang2019chemprop}, and information
flows through the hierarchy so that each level is informed by the
others.
Figure~\ref{fig:architecture}b illustrates this hierarchical message
passing on a concrete example, showing how atom-level, motif-level,
and monomer-level representations interact for a multi-component
polymer with its initiator and terminator fragments.

% ---------------------------------------------------------------------------
% Figure 2: Architecture, parity plots, and ablation
% ---------------------------------------------------------------------------

Three physically motivated design principles distinguish \modelname{}
from prior flat-graph approaches.
The first is stochastic edge weighting.
Within each monomer fragment, the covalent bonds are deterministic, and
the message passing at the atom and motif levels proceeds with uniform
weights.
However, at the monomer graph level, the connectivity between
fragments reflects the stochastic nature of polymerization: which
monomers are adjacent in a given chain depends on reaction kinetics
and thermodynamics.
\modelname{} encodes this distinction by selectively applying \grins{}-derived
bond probabilities as edge weights at the monomer graph
level, modulating how strongly each fragment node integrates
information from its neighbors.

The second principle is composition-aware aggregation.
After multi-resolution pooling condenses each fragment's node
embeddings into a single fixed-size vector, the fragment
representations are combined into a polymer-level embedding using
mole fractions as aggregation weights,
$\mathbf{h}_{\mathrm{polymer}} = \sum_i w_i \cdot \mathbf{h}_i$ (Fig.~\ref{fig:architecture}a, Prediction Head).
This design mirrors the linear mixing rules that govern many polymer
properties~\cite{fox1952tg} and encodes the composition directly into the
learned representation rather than providing it as auxiliary metadata.

The third principle is hierarchical context integration through a root
vector.
A root vector summarizes the global chemistry information encoded
across the hierarchical levels and is injected into the prediction head
alongside the composition-aggregated polymer embedding
(Fig.~\ref{fig:architecture}a, Root Vector).
This combines the local chemical detail captured by fragment-level
pooling with a global structural context that reflects how functional
groups are organized across the full molecular hierarchy.

In addition to composition and connectivity, the model incorporates
molecular weight through a dedicated transformation in the prediction
head (Fig.~\ref{fig:architecture}a). When the degree of polymerization
$N$ is specified in the \grins{} string, the polymer-level embedding
receives an additive chain-end correction that decays as $1/N$ and
vanishes in the infinite-chain limit, the dependence underlying the
Fox--Flory relation~\cite{fox_flory1950,rubinstein2003polymer}
(Methods, Eq.~\ref{eq:mw}). For properties that follow power laws in
chain length, the prediction head additionally receives $\log N$. When
the full molecular weight distribution is available, \modelname{}
instead incorporates it through a distribution-aware
encoding~\cite{hu2025mw} that captures dispersity and higher-order
moments.

% ---------------------------------------------------------------------------
\subsection*{Prediction performance and comparison with baselines}
\label{sec:results-prediction}
% ---------------------------------------------------------------------------

We evaluated \modelname{} on two fundamental thermophysical properties,
the glass transition temperature $T_g$ and the mass density $\rho$,
 calculated using molecular dynamics simulations using \grins{}-generated
chain ensembles under identical thermodynamic conditions.
To assess the advantages of a polymer-native, hierarchical
architecture, we compared \modelname{} against four baselines that
collectively represent the dominant paradigms in current polymer
property prediction: fingerprint-based machine learning, pretrained
language models, and graph neural networks
(Table~\ref{tab:capability}).
These include random forest (RF) regression paired with
extended-connectivity fingerprints~\cite{rogers2010ecfp} in two
composition-handling variants (RF~+~ECFP with composition as
auxiliary features, RF~+~wFP with composition-weighted fingerprint
averaging), polyBERT~\cite{kuenneth2023polybert} with
composition-weighted PSMILES embedding averaging, and Polymer
Chemprop~\cite{aldeghi2022wdmpnn}, a weighted directed
message-passing neural network (wD-MPNN) that operates on a flat
molecular graph.
All baselines were re-implemented on our dataset and evaluated under
identical five-fold cross-validation splits.

%% TABLE 1: Capability comparison
\begin{table}[ht]
  \centering
  \caption{\textbf{Methodological comparison across approaches.}}
  \label{tab:capability}
  \resizebox{\textwidth}{!}{%
  \begin{tabular}{llcccccc}
    \toprule
    \multicolumn{1}{c}{Method}
      & \multicolumn{1}{c}{Representation}
      & \multicolumn{1}{c}{\begin{tabular}[c]{@{}c@{}}Polymer-\\native\end{tabular}}
      & \multicolumn{1}{c}{\begin{tabular}[c]{@{}c@{}}Multi-\\comp.\end{tabular}}
      & \multicolumn{1}{c}{\begin{tabular}[c]{@{}c@{}}Mol.\\frac.\end{tabular}}
      & \multicolumn{1}{c}{Conn.}
      & \multicolumn{1}{c}{MW}
      & \multicolumn{1}{c}{Gen.} \\
    \midrule
    RF + ECFP$^\dagger$
      & Monomer ECFP$^\ddagger$
      & \xmark & $\sim$ & $\sim$ & \xmark & \xmark & \xmark \\
    RF + wFP$^\dagger$
      & Monomer ECFP$^\ddagger$
      & \xmark & $\sim$ & $\sim$ & \xmark & \xmark & \xmark \\
    polyBERT$^\dagger$
      & Monomer PSMILES$^\ddagger$
      & $\sim$ & $\sim$ & $\sim$ & \xmark & \xmark & \xmark \\
    Polymer Chemprop$^\dagger$
      & Flat molecular graph
      & \cmark & \cmark & \cmark & \pmark & \pmark & \xmark \\
    \addlinespace
    \textbf{\modelname{} (This work)}
      & \textbf{Hierarchical \grins{} graph}
      & \cmark & \cmark & \cmark & \cmark & \cmark & \cmark \\
    \bottomrule
  \end{tabular}%
  }
  \\[4pt]
  \raggedright
  {\footnotesize
    \cmark{} = natively supported;
    \pmark{} = limited native support;
    $\sim$ = supported via our adaptation;
    \xmark{} = not supported.
    Multi-comp.\ = multi-component polymers;
    Mol. frac. = mole fractions;
    Conn.\ = inter-monomer connectivity; \\
    MW = molecular weight;
    Gen.\ = generative design. \\
    $^\dagger$ Re-implemented on our dataset. \\
    $^\ddagger$ Adapted for multi-component polymers:
    RF + ECFP uses composition as auxiliary features;
    RF + wFP uses composition-weighted fingerprint averaging;
    polyBERT uses composition-weighted embedding averaging.
  }
\end{table}

Table~\ref{tab:capability} compares the methodological capabilities
of these approaches.
The fingerprint and language model baselines operate on individual
monomer representations and have no native mechanism for encoding how
multiple monomers combine; to enable a fair comparison on our
multi-component dataset, we adapted them as indicated in the table
footnotes.
Polymer Chemprop natively accepts multi-component polymers and
encodes mole fractions and inter-monomer connectivity through its
bond weighting scheme, though it relies on pre-fixed hypothetical
bond weights assigned by the idealized architecture type, such as random,
block, or alternating, rather than deriving connectivity from the
actual polymer specification.
Only \modelname{} provides a fully polymer-native representation that
encodes mole fractions, stochastic inter-monomer bond topology, and
molecular weight directly within a hierarchical architecture, while
additionally supporting generative molecular design through its
shared variational latent space.

The performance comparison in Table~\ref{tab:baselines} reveals
a clear progression from monomer to polymer
representations.
\modelname{} achieves the highest accuracy in both properties:
$R^2 = 0.803 \pm 0.147$ for $T_g$ and
$R^2 = 0.945 \pm 0.024$ for density
(Table~\ref{tab:baselines}; 
a representative fold is shown in Fig.~\ref{fig:architecture}c).
Among the baselines, the comparison between RF~+~ECFP and RF~+~wFP
is instructive: these methods differ only in how composition enters
the representation, yet composition-weighted averaging nearly doubles
the $T_g$ $R^2$ from 0.36 to 0.70, demonstrating that encoding
composition within the molecular representation is essential for
multi-component systems.
polyBERT achieves comparable performance to the weighted fingerprint
baseline despite access to a much larger pretraining corpus,
suggesting that representations learned by monomer-level language
models do not transfer effectively to polymer-level property
prediction because the pretraining objective captures intra-monomer
chemical patterns but not inter-monomer interactions or
composition-dependent property trends.
Polymer Chemprop achieves strong density predictions
($R^2 = 0.884$), as density depends mainly on atomic packing and
van der Waals volume, which are well captured by atom-level graph
representations.
However, its $T_g$ predictions are notably unstable, with a
cross-fold standard deviation of 0.427, indicating that some folds
essentially fail; this reflects the greater sensitivity of $T_g$ to
cooperative segmental dynamics and inter-monomer interactions that
require explicit hierarchical encoding.

%% TABLE 2: Performance comparison
\begin{table}[ht]
  \centering
  \caption{\textbf{Prediction performance across methods.}
    Five-fold cross-validation; mean $\pm$ s.d.\ across folds.
    All methods trained on identical splits.
  }
  \label{tab:baselines}
  \resizebox{\textwidth}{!}{%
  \begin{tabular}{lcccccc}
    \toprule
    & \multicolumn{3}{c}{$T_g$} & \multicolumn{3}{c}{Density} \\
    \cmidrule(lr){2-4} \cmidrule(lr){5-7}
    \multicolumn{1}{c}{Method}
      & $R^2$ & RMSE (K) & MAE (K)
      & $R^2$ & RMSE (g\,cm$^{-3}$) & MAE (g\,cm$^{-3}$) \\
    \midrule
    RF + ECFP$^\dagger$
      & $0.362 \pm 0.540$ & $9.912 \pm 2.537$ & $7.744 \pm 1.930$
      & $0.532 \pm 0.224$ & $0.0202 \pm 0.0075$ & $0.0149 \pm 0.0056$ \\
    RF + wFP$^\dagger$
      & $0.695 \pm 0.137$ & $7.548 \pm 1.778$ & $5.435 \pm 1.387$
      & $0.784 \pm 0.064$ & $0.0135 \pm 0.0033$ & $0.0100 \pm 0.0025$ \\
    polyBERT$^\dagger$
      & $\underline{0.759 \pm 0.201}$ & $\underline{6.480 \pm 2.421}$ & $\underline{4.628 \pm 1.453}$
      & $0.850 \pm 0.145$ & $0.0106 \pm 0.0050$ & $0.0070 \pm 0.0024$ \\
    Polymer Chemprop$^\dagger$
      & $0.618 \pm 0.427$ & $7.682 \pm 3.972$ & $5.832 \pm 3.050$
      & $\underline{0.884 \pm 0.054}$ & $\underline{0.0099 \pm 0.0031}$ & $\underline{0.0069 \pm 0.0023}$ \\
    \addlinespace
    \modelname{} (This work)
      & $\mathbf{0.803 \pm 0.147}$ & $\mathbf{5.900 \pm 2.454}$ & $\mathbf{4.240 \pm 1.579}$
      & $\mathbf{0.945 \pm 0.024}$ & $\mathbf{0.0068 \pm 0.0025}$ & $\mathbf{0.0052 \pm 0.0021}$ \\
    \bottomrule
  \end{tabular}%
  }
  \\[4pt]
  \raggedright
  {\footnotesize
    $^\dagger$ Re-implemented on our dataset with composition adaptation where applicable. \\
    \textbf{Bold} = best; \underline{underline} = second best.
  }
\end{table}

\modelname{} resolves these limitations and achieves the strongest
results on both properties.
For $T_g$, it reaches an MAE of 4.2~K with the lowest cross-fold
variance among the graph-based methods.
For density, it achieves an MAE of 0.005~g \, cm$^{-3}$ with
$R^2 = 0.945$, surpassing the next-best method by more than 6 percentage
points.
This error approaches the intrinsic uncertainty of the molecular
dynamics simulations used to generate the target values, suggesting
that \modelname{} extracts nearly all the learnable information from
the representation \grins{} for this property.
The tight clustering along the diagonal in the parity plots
(Fig.~\ref{fig:architecture}c) visually confirms this quantitative assessment, 
with density predictions showing a particularly narrow
scatter across training, validation, and test partitions.
 
To understand which architectural components drive this performance,
we conducted an ablation study using $T_g$ as the representative
property, as its sensitivity to cooperative segmental dynamics
amplifies the effect of each design choice
(Fig.~\ref{fig:architecture}d).
Each variant removes exactly one component from the complete model
($R^2 = 0.803$), keeping all others intact.
Removing composition-aware aggregation collapses prediction to
$R^2 = 0.291$, as expected for multi-component systems where
composition is a primary determinant of properties.
More revealing is that the remaining three components each
independently improve accuracy.
Stochastic edge weighting encodes which monomers are likely
neighbors in the chain, capturing sequence-level conformational
effects that cannot be recovered from monomer chemistry alone.
Root vector injection modulates the prediction head with a global
summary of the hierarchical encoding, providing a top-down structural
context that guides property estimation.
Root vector concatenation enriches the polymer embedding with this
same hierarchical information after composition-aware aggregation,
supplying complementary chemical detail at the feature level.
Each addresses a distinct aspect of polymer structure that
flat-graph approaches cannot represent.

% ---------------------------------------------------------------------------
\subsection*{Generative design of sustainable polymer alternatives}
\label{sec:results-pfas}
% ---------------------------------------------------------------------------

\begin{figure}[!ht]
  \centering
  \includegraphics[width=\textwidth]{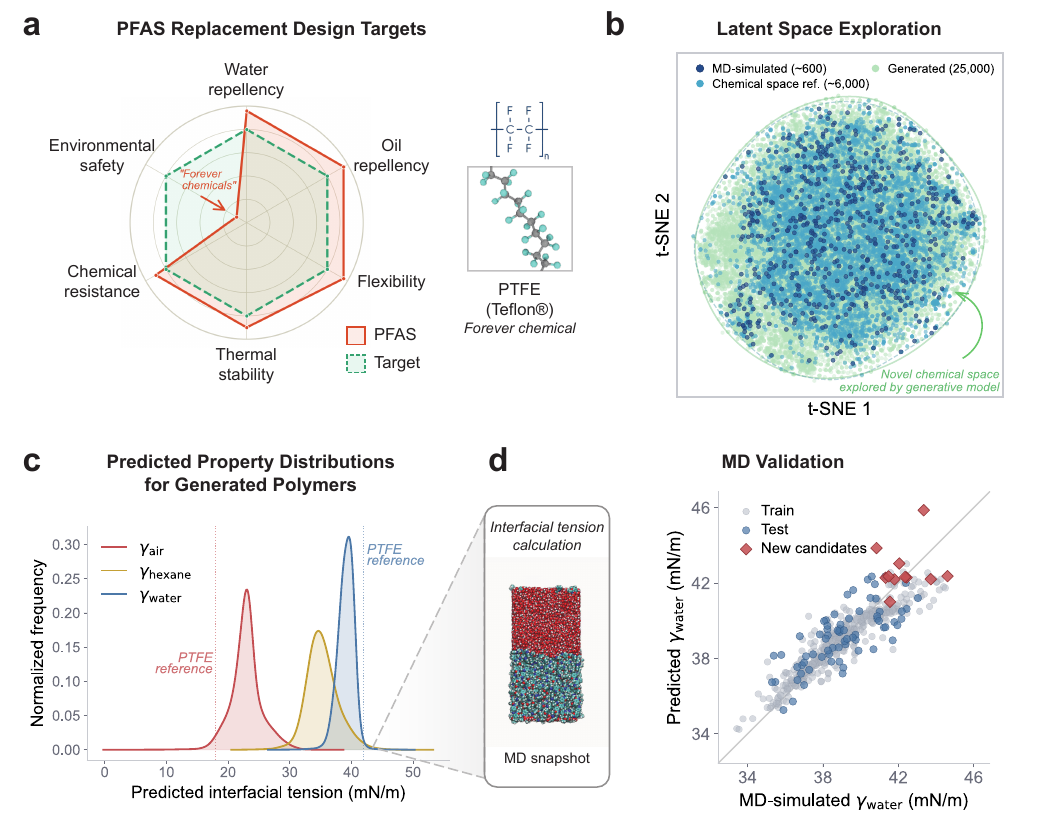}
  \caption{%
    \textbf{Generative design of PFAS-free polymer alternatives.}
    \textbf{a},~Multi-objective design targets for PFAS replacement.
    The radar chart compares the property profile of PFAS (red) with
    the desired target (green dashed) across water repellency, oil
    repellency, flexibility, chemical resistance, thermal stability,
    and environmental safety.
    PTFE (Teflon\textsuperscript{\textregistered}) is shown as the
    reference fluorinated polymer.
    \textbf{b},~t-SNE projection of the learned latent space.
    The reference chemical space comprises approximately 6{,}000
    unlabeled polymers (light blue), of which approximately 600 are
    labeled through MD simulation (dark blue).
    The generative model samples 25{,}000 novel candidates (green)
    that fill and expand beyond the reference space.
    \textbf{c},~Predicted interfacial tension distributions for the
    generated polymers at polymer--air, polymer--hexane, and
    polymer--water interfaces.
    Dashed lines indicate PTFE reference values.
    \textbf{d},~MD validation of top-ranked candidates.
    The parity plot compares predicted and MD-simulated
    $\gamma_{\mathrm{water}}$ for training data (gray), test data
    (blue), and newly validated candidates (red).
    Inset: MD simulation snapshot of the interfacial tension
    calculation.
  }
  \label{fig:pfas}
\end{figure}

To demonstrate the full discovery capability of \modelname{}, we
applied the generative design pathway to a pressing real-world
challenge: identifying sustainable replacements for per- and
polyfluoroalkyl substances (PFAS).
PFAS, exemplified by polytetrafluoroethylene (PTFE), are widely used
for their exceptional water and oil repellency, chemical resistance,
and thermal stability, yet their extreme environmental persistence
has led to tightening of regulatory restrictions
worldwide~\cite{pfas_regulation}.
The design objective is to discover polymers that approach the
surface-energy performance of PFAS while eliminating the fluorinated
chemistries responsible for their environmental accumulation
(Fig.~\ref{fig:pfas}a).
We quantify surface energy performance through interfacial tensions
at polymer--air ($\gamma_{\mathrm{air}}$),
polymer--hexane ($\gamma_{\mathrm{hexane}}$) and
polymer--water ($\gamma_{\mathrm{water}}$) interfaces, which
collectively characterize hydrophobicity, oleophobicity and surface
energy.

% ---------------------------------------------------------------------------
% Figure 3: Generative design for PFAS replacement
% ---------------------------------------------------------------------------

We began by constructing a chemical reference space from
approximately 6{,}000 polymers encoded as \grins{} strings
(Fig.~\ref{fig:pfas}b, light blue).
From this pool, approximately 600 were selected and labeled through
the automated MD pipeline described in Methods, resulting in interfacial
tension values computed under identical simulation conditions
(Fig.~\ref{fig:pfas}b, dark blue).
The t-SNE projection of the learned latent space confirms that these
labeled polymers cover the reference chemical space, providing a
representative training set despite their modest size.

Using the shared variational latent space of \modelname{}, we then
sampled 25{,}000 novel polymer candidates
(Fig.~\ref{fig:pfas}b, green).
These generated polymers not only fill gaps within the existing
chemical space but also expand into previously unexplored regions,
demonstrating that the generative model discovers structurally
diverse candidates beyond the training distribution.
The trained predictor, using only 600 labeled polymers,
selected all 25{,}000 candidates for their interfacial tension
profiles (Fig.~\ref{fig:pfas}c).
The predicted distributions reveal a population of generated
polymers that exhibit simultaneously high $\gamma_{\mathrm{water}}$
and low $\gamma_{\mathrm{air}}$, which match the surface-energy
signature of PTFE while avoiding fluorinated building blocks.

To confirm that these predictions reflect genuine material
properties rather than artifacts of extrapolation, we independently
validated the top-ranked candidates through MD
simulations using \grins{}-generated chain ensembles
(Fig.~\ref{fig:pfas}d).
The parity plot comparing the predicted and MD-simulated
$\gamma_{\mathrm{water}}$ shows that the newly validated candidates
(red) fall along the same diagonal as the original training and test
data (gray and blue), confirming that \modelname{} generalizes
reliably to new chemistries outside the training set.
This validation step closes the feedback loop introduced in
Fig.~\ref{fig:workflow}: The generated candidates are confirmed by
 simulation based on physics using the same representation \grins{} that 
drives both the neural network and the MD pipeline, ensuring
end-to-end self-consistency.

The resulting PFAS-free candidate structures are chemically diverse and feature
unconventional functional group combinations, including aromatic-ester
backbones, heterocyclic units, and sterically hindered side groups,
which would be difficult to arrive at through human intuition or
conventional structure--property heuristics alone.
This highlights a key strength of the generative approach: by
operating in a learned latent space rather than enumerating known
chemistries, \modelname{} explores regions of polymer space that lie
outside established design rules, identifying candidates whose
surface-energy performance arises from subtle interplay between
backbone rigidity, side-chain polarity, and intermolecular packing.
More broadly, this case study illustrates how the \modelname{}
framework can accelerate the discovery of functional polymer
materials in application domains where the design space is vast,
experimental iteration is costly, and traditional material selection
is increasingly constrained by environmental demands.

% ===========================================================================
% EXTENDED DATA FIGURES
% ===========================================================================
%
%
% ===========================================================================

% ============================================================================
% Discussion
% ============================================================================

\section*{Discussion}

In this work, we have presented \modelname{}, a polymer-native hierarchical
 AI framework that operates on complete polymer
descriptions rather than monomer-level proxies.
\modelname{} encodes the polymer structure in three levels of resolution, from
atoms through functional group motifs to monomer-level connectivity,
and captures composition, stochastic bonding, and molecular weight
natively within a single end-to-end architecture built on the
\grins{} representation.
By coupling this hierarchical encoder with both a multi-target
property predictor and a variational generative model, \modelname{}
establishes a complete workflow for polymer discovery, from
formulation data to validated material candidates.

A notable consequence of this hierarchical and physically informed
design is data efficiency.
\modelname{} achieves the highest accuracy of prediction for $T_g$
and density, and identifies validated PFAS-free candidates, in both
cases from modestly sized training sets.
This is possible because the architecture encodes domain knowledge
directly into its structure instead of relying on large-scale
pretraining to implicitly learn polymer physics from data.
For multi-component polymer systems, where labeled data remain
scarce and expensive to obtain through either experiment or
simulation, this data efficiency is a practical advantage. 

The design principles validated here extend beyond \modelname{}
itself.
The strategies of encoding composition through weighted aggregation,
incorporating connectivity through probabilistic edge weights, and
integrating hierarchical context through root vectors are general
and can be adopted by other graph neural network architectures for
polymer property prediction.
The framework further generalizes to any scalar polymer property for
which training data are available, including mechanical, rheological,
transport, and dielectric properties, offering a broadly applicable
foundation for polymer-native ML.

In the future, expanding the training data to include broader
polymer families and additional properties would strengthen both
predictive accuracy and generative diversity.
Integration with experimental validation, where \grins{}-generated
candidates are synthesized and characterized in the laboratory,
would close the complete design loop from the computation to the physical
material.
The PFAS case study demonstrates that a polymer-native AI framework
can address real-world formulation challenges with practical impact.
We anticipate that the principles established here, polymer-native
representation, hierarchical physical encoding, and self-consistent
validation through \grins{}, will inform the next generation of
polymer AI models, ultimately accelerating the design of polymer
materials for applications ranging from sustainable packaging to
advanced energy storage.

% ============================================================================
% Methods Section
% ============================================================================

\section*{Methods}

% ---------------------------------------------------------------------------
\subsection*{Problem formulation}
% ---------------------------------------------------------------------------

We formulate polymer property prediction as multi-target regression
from molecular structure.
Given a polymer defined by a set of $K$ monomer repeat units with
SMILES representations $\{s_1, \ldots, s_K\}$, mole fractions
$\{w_1, \ldots, w_K\}$ satisfying $\sum_{i=1}^{K} w_i = 1$,
inter-monomer bond connectivity rules
$\mathcal{B} = \{(i, j, p_{ij}, p_{ji})\}$ specifying stochastic
bond probabilities between attachment sites, and a molecular weight 
distribution $\text{MW} \sim \mathcal{D}(\boldsymbol{\theta})$
where $\mathcal{D}$ may be Gaussian, uniform, Poisson, Flory--Schulz,
Schulz--Zimm, or log-normal, the
task is to predict a vector of continuous material properties
$\mathbf{y} \in \mathbb{R}^{P}$ (e.g., glass transition temperature
$T_g$, density $\rho$, interfacial tension $\gamma$).
The model handles linear and branched polymers with two or more
monomer types, variable compositions, and multiple attachment
topologies, and supports training with incomplete property labels via
NaN-aware loss computation.

% ---------------------------------------------------------------------------
\subsection*{Polymer representation}
\label{sec:methods-representation}
% ---------------------------------------------------------------------------

Each polymer is encoded as a pipe-delimited string derived from the
\grins{} representation~\cite{grins}, which builds on the BigSMILES
line notation for stochastic
macromolecules~\cite{bigsmiles2019} and its generative
extension~\cite{gbigsmiles2024}:
\begin{equation}
  \underbrace{s_1.s_2.\ldots.s_K}_{\text{monomer SMILES}}
  \;\Big|\; \underbrace{w_1 | w_2 | \ldots | w_K}_{\text{mole fractions}}
  \;\Big|\; \underbrace{\langle i\text{-}j : p_{ij} : p_{ji} \rangle \ldots}_{\text{bond rules}}
  \;\Big|\; \underbrace{\mathcal{D}(\boldsymbol{\theta})}_{\text{MW distribution}}
\end{equation}

Monomer SMILES contain wildcard attachment sites
$[\ast\!:\!n]$ that define inter-monomer bonding points.
The bond rules specify, for each pair of attachment sites $(i,j)$,
the directional probabilities $p_{ij}$ and $p_{ji}$ with which those
sites are bonded in the polymer ensemble.

These bond probabilities and mole fractions are not analytically
assumed from idealized architectures (e.g., setting $p = 0.5$ for
random polymers as in prior
work~\cite{aldeghi2022wdmpnn}).
Instead, they are derived from realistic chain ensembles generated by
\grins{}, which produces statistically representative populations of
polymer chains that respect the specified composition, connectivity,
and molecular weight distribution.
The same chain ensembles serve as input to MD simulations,
providing a self-consistent framework in which the input to the AI
model is grounded in the same physical representation used for
atomistic validation.

The molecular weight field specifies a parametric distribution from a
supported family (Gaussian, uniform, Poisson, Flory--Schulz,
Schulz--Zimm, or log-normal), or alternatively a histogram for
empirical distributions, from which chain lengths are sampled during
ensemble generation.

% ---------------------------------------------------------------------------
\subsection*{Hierarchical graph construction}
\label{sec:methods-graph}
% ---------------------------------------------------------------------------

\paragraph{Fragment decomposition.}
The \grins{} string is parsed into its constituent monomer fragments.
Wildcard atoms are removed and the resulting molecular graph is
split into $K$ disconnected monomer fragments.
Per-atom weight fractions (derived from mole fractions) and
attachment-site metadata are retained as node attributes.

\paragraph{Inter-monomer bond feature extraction.}
Bond features for inter-monomer connections, which do not correspond
to permanent covalent bonds in any single fragment, are extracted via
a duplication method: the core molecule is duplicated, a temporary
bond is inserted between attachment sites across the original and
copy, bond features are extracted, and the temporary bond is removed.
This produces chemically consistent bond descriptors (bond type,
conjugation, ring membership, stereo) together with the directional
connection probabilities $(p_{ij}, p_{ji})$ that modulate the message
passing at the monomer graph level.

\paragraph{Functional group decomposition.}
Each monomer fragment independently undergoes functional group
decomposition~\cite{han2025}:
\begin{enumerate}
  \item \textit{Cluster identification}: atoms are grouped into
    elementary clusters defined by individual bonds and simple rings
    (determined by the smallest set of smallest rings).
  \item \textit{Motif pooling}: clusters are merged into higher-level
    functional group motifs using a pre-computed vocabulary of
    chemically meaningful substructures (ring systems, ether linkages,
    ester groups) derived from the training data.
  \item \textit{Graph construction}: a graph is built over the
    resulting motifs, defining parent--child relationships through a
    maximum spanning approach.
  \item \textit{Traversal ordering}: depth-first search labels graph
    edges with traversal order, used for positional encoding during
    message passing.
\end{enumerate}
The result is a two-view representation per fragment: an atom-level
molecular graph and a motif-level graph, linked by a
cluster-to-atom mapping matrix $\mathbf{C}$ that records which atoms
belong to which motif.

\paragraph{Polymer-level graph assembly.}
Fragment-level motif graphs are merged into a unified polymer graph
with appropriate index offsets.
Inter-monomer connection metadata, including bond features,
directional weights, and polymer-level attributes (degree of
polymerization, composition), are attached to graph nodes, enabling
polymer-aware message passing.

% ---------------------------------------------------------------------------
\subsection*{Three-level hierarchical encoder}
\label{sec:methods-encoder}
% ---------------------------------------------------------------------------

The encoder processes the polymer graph through three levels of
message passing, each operating on a different structural resolution.
All three levels use the same underlying mechanism, a directed
message-passing neural network
(D-MPNN)~\cite{yang2019chemprop} with LSTM-based~\cite{hochreiter1997lstm}
recurrent message updates, but differ in the graph topology on which they
operate and whether stochastic edge weights are applied.

Information flows bottom-up through the hierarchy: the atom graph
(Level~3) is encoded first, its embeddings feed into the motif graph
(Level~2), and the motif representations in turn feed into the
monomer graph (Level~1).

\paragraph{Directed message passing with stochastic edge weighting.}
At each level $\ell$, the edge-centered hidden states
$\mathbf{h}_{vu}^{(\ell,t)}$ are updated over $D_\ell$ iterations:
\begin{equation}
  \mathbf{m}_{vu}^{(\ell,t)} = \sum_{k \in \mathcal{N}(v) \setminus u} \mathbf{h}_{kv}^{(\ell,t-1)}
\end{equation}
\begin{equation}
  \mathbf{h}_{vu}^{(\ell,t)} = \mathrm{LSTM}\!\left(\mathbf{h}_{vu}^{(\ell,t-1)},\; \mathbf{m}_{vu}^{(\ell,t)}\right)
\end{equation}
After $D_\ell$ iterations, node representations are obtained by
aggregating incoming edge messages:
\begin{equation}
  \mathbf{h}_v^{(\ell)} = \tau\!\left(\mathbf{W}_o^{(\ell)} \left[\mathbf{x}_v^{(\ell)} \;\|\; \sum_{k \in \mathcal{N}(v)} \alpha_{kv}^{(\ell)} \, \mathbf{h}_{kv}^{(\ell,D_\ell)} \right]\right)
\end{equation}
where $\mathbf{x}_v^{(\ell)}$ is the input feature vector at level
$\ell$, $\|$ denotes concatenation, $\tau$ is a nonlinear activation,
and $\alpha_{kv}^{(\ell)} \in [0,1]$ is an edge-level aggregation
weight that modulates the contribution of the message from node $k$
to node $v$.

The edge weights $\alpha_{kv}^{(\ell)}$ encode the central physical
insight of the model.
At the level of the monomer graph (Level~1), the edges represent inter-monomer
connections whose presence is stochastic, so
$\alpha_{kv}^{(1)} = p_{kv}$ where $p_{kv}$ is the probability of bonds derived from \grins{}
 between fragments $k$ and $v$.
At the level of the motif graph (Level~2), the topology of the graph is fixed for a
given molecular structure, so $\alpha_{kv}^{(2)} = 1$.
At the atom graph level (Level~3), all bonds within a monomer
fragment are deterministic covalent bonds, so
$\alpha_{kv}^{(3)} = 1$ (standard unweighted message passing).

This selective application of stochastic edge weights reflects the
physical reality that structural uncertainty in polymers resides
primarily at inter-monomer connections, where the stochastic
polymerization process determines which monomers are adjacent.

\paragraph{Level 1: Monomer graph encoding (coarsest).}
The monomer graph encoder operates on the connectivity among monomer
fragments, with edges annotated by \grins{}-derived bond
probabilities.
Input features for each monomer node are formed by concatenating a
learned vocabulary embedding $\mathbf{E}_c(v)$ (encoding the motif
identity) with the motif graph representation from Level~2:
\begin{equation}
  \mathbf{x}_v^{(1)} = \tau\!\left(\mathbf{W}_c \left[\mathbf{E}_c(v) \;\|\; \mathbf{h}_v^{(2)}\right]\right)
\end{equation}
This level captures long-range inter-monomer interactions across the
polymer structure, modulated by stochastic edge weights.

\paragraph{Level 2: Motif graph encoding.}
The motif graph encoder operates on the functional group graph within
each fragment.
The input features for each node of the motif $v$ are formed by concatenating a
learned vocabulary embedding $\mathbf{E}_i(v)$ (encoding the identity of the motif
) with the sum of the atom embeddings belonging to that motif:
\begin{equation}
  \mathbf{x}_v^{(2)} = \tau\!\left(\mathbf{W}_i \left[\mathbf{E}_i(v) \;\|\; \sum_{a \in \mathcal{C}(v)} \mathbf{h}_a^{\mathrm{atom}}\right]\right)
\end{equation}
where $\mathcal{C}(v)$ denotes the set of atoms in motif $v$ (from
the cluster-to-atom mapping $\mathbf{C}$) and $\mathbf{W}_i$ is a
learned linear projection.
Edge features include positional encodings from the depth-first
traversal order.
After $D_2$ iterations, the output is motif-level embeddings
$\mathbf{h}_v^{(2)} \in \mathbb{R}^d$.

\paragraph{Level 3: Atom graph encoding (finest).}
The atom graph encoder operates on the molecular graph of the covalent
bonds within each monomer fragment.
The features of the input node $\mathbf{x}_v^{(3)}$ are one-hot encodings in
a vocabulary of pairs of (element symbol, formal charge), which capture the identity of the atom.
Edge features are one-hot encodings of bond type (single, double,
triple, aromatic) combined with positional encodings from the
traversal order.
After $D_3$ iterations of message passing, the output is atom
embeddings $\mathbf{h}_v^{\mathrm{atom}} \in \mathbb{R}^{d}$ for all
atoms in the batch.

\paragraph{Root vector.}
For each fragment in the batch, a root embedding
$\mathbf{h}^{\mathrm{root}}$ is computed by aggregating the incoming
motif-level edge messages at the designated root node of the motif
graph:
\begin{equation}
  \mathbf{h}^{\mathrm{root}} = \tanh\!\left(\mathbf{W}_r \left[\mathbf{x}_{r}^{(1)} \;\|\; \sum_{k \in \mathcal{N}(r)} \mathbf{h}_{kr}^{(1, D_1)}\right]\right)
\end{equation}
where $r$ is the root node, $\mathbf{x}_{r}^{(1)}$ is the root's
input feature at Level~1, $\mathbf{h}_{kr}^{(1, D_1)}$ are the final
edge-state messages arriving at $r$ after $D_1$ iterations, and
$\mathbf{W}_r \in \mathbb{R}^{d \times 2d}$.
The root vector combines a node's local motif identity with the
global message-passing context, providing a summary of the entire
hierarchical structure of each monomer fragment.

% ---------------------------------------------------------------------------
\subsection*{Fragment-level embedding}
\label{sec:methods-fragment-embedding}
% ---------------------------------------------------------------------------

The hierarchical encoder produces per-node representations at
multiple levels.
To obtain a fixed-size embedding per fragment, we use an
attention-based pooling module.
For each fragment, the pooling proceeds in two branches:
\begin{enumerate}
  \item A node-level MLP produces per-node embeddings
    $\mathbf{h}_v^{(a)}$.
  \item A second MLP produces pairwise embeddings
    $\mathbf{h}_v^{(b)}$, which are processed through multi-head
    self-attention~\cite{vaswani2017attention} to capture inter-node
    interactions.
    The attention output is element-wise multiplied with
    $\mathbf{h}_v^{(b)}$ (gating), then concatenated with
    $\mathbf{h}_v^{(a)}$ and projected.
\end{enumerate}
The resulting per-node representations are aggregated into a single
fragment embedding via multi-resolution pooling, which concatenates
four complementary strategies and projects to the embedding dimension
$d_e$:
\begin{equation}
  \mathbf{h}^{\mathrm{frag}} = \mathbf{W}_{\mathrm{pool}} \left[
    \underbrace{\frac{1}{|\mathcal{V}|}\sum_{v} \mathbf{h}_v}_{\text{mean}} \;\Big\|\;
    \underbrace{\max_{v}\, \mathbf{h}_v}_{\text{max}} \;\Big\|\;
    \underbrace{\sum_{v} \sigma(\mathbf{g}_v) \odot \tanh(\mathbf{f}_v)}_{\text{gated}} \;\Big\|\;
    \underbrace{\sum_{v} a_v \, \mathbf{h}_v}_{\text{attention}}
  \right]
\end{equation}
where $\sigma(\mathbf{g}_v)$ and $\tanh(\mathbf{f}_v)$ are learned
gate and filter functions,
$a_v = \mathrm{softmax}_v(\mathbf{q}(\bar{\mathbf{h}})^\top \mathbf{k}(\mathbf{h}_v))$
are attention weights with query $\mathbf{q}$ derived from the mean
embedding $\bar{\mathbf{h}}$ and per-node keys $\mathbf{k}$, and
$\mathbf{W}_{\mathrm{pool}} \in \mathbb{R}^{d_e \times 4d_e}$.
This multi-resolution strategy addresses the well-known limitation
that different material properties may depend on different
statistical aspects of the molecular representation.

% ---------------------------------------------------------------------------
\subsection*{Composition-aware aggregation}
\label{sec:methods-adapter}
% ---------------------------------------------------------------------------

The fragment-level embedding produces one vector per monomer
fragment, whereas the prediction head requires one embedding per
polymer.
The composition-aware aggregation bridges this gap through
mole-fraction-weighted summation:
\begin{equation}
  \mathbf{h}^{\mathrm{poly}} = \sum_{i=1}^{K} w_i \, \mathbf{h}_i^{\mathrm{frag}}
  \label{eq:weighted-agg}
\end{equation}
where $w_i$ is the normalized mole fraction of monomer $i$ and $K$ is
the number of distinct monomers.
This formulation is physically motivated by the observation that many
polymer properties follow approximate linear mixing rules, for
example, the Fox equation~\cite{fox1952tg} for the glass transition
temperature $1/T_g = \sum_i w_i / T_{g,i}$, which in a 
Taylor expansion of the first-order is linear in composition fractions.
Composition-weighted aggregation encodes this inductive bias directly
into the architecture, in contrast to approaches that treat
composition as an auxiliary input or metadata feature.

\paragraph{Root vector integration.}
The root embeddings $\mathbf{h}^{\mathrm{root}}$ from the
hierarchical encoder are separately aggregated to the polymer level
using the same composition weighting (Eq.~\ref{eq:weighted-agg}),
projected to match the embedding dimension of the fragment via a learned
linear map
$\mathbf{W}_{\mathrm{root}} \in \mathbb{R}^{d_e \times d}$, and
concatenated with the embedding of the polymer:
\begin{equation}
  \tilde{\mathbf{h}}^{\mathrm{poly}} = \mathbf{W}_{\mathrm{adapt}} \left[
    \mathbf{h}^{\mathrm{poly}} \;\|\; \mathbf{W}_{\mathrm{root}}\, \mathbf{h}^{\mathrm{root,poly}}
  \right]
\end{equation}
where
$\mathbf{W}_{\mathrm{adapt}} \in \mathbb{R}^{d_e \times 2d_e}$.
This integration combines local chemical detail from the fragment
embeddings with a global hierarchical context from the root vectors.
Ablation experiments confirm that both components contribute to
the accuracy of the prediction and that their combination outperforms either one
alone (Fig.~\ref{fig:architecture}d).

% ---------------------------------------------------------------------------
\subsection*{Molecular weight transformation}
\label{sec:methods-mw}
% ---------------------------------------------------------------------------
When the degree of polymerization $N$ is specified in the \grins{} string,
the polymer-level embedding receives an additive chain-end correction:
\begin{equation}
  \label{eq:mw}
  \tilde{\mathbf{h}}^{\mathrm{poly}}_{\mathrm{MW}}
  = \tilde{\mathbf{h}}^{\mathrm{poly}}
  + \frac{N_0}{N + N_0}\,\mathbf{W}_{\mathrm{end}}\,
    \tilde{\mathbf{h}}^{\mathrm{poly}}
\end{equation}
The first term is the embedding of the infinitely long chain. The second
term accounts for chain ends. Each chain of $N$ repeat units carries two
ends, so the chain-end concentration scales as $1/N$, and to leading order
chain ends shift bulk properties in proportion to their
concentration~\cite{fox_flory1950,rubinstein2003polymer}. For $N \gg N_0$
the correction decays as $N_0/N$ and vanishes in the infinite-chain limit,
so that with a linear readout a predicted property takes the Fox--Flory
form $y \simeq y_\infty - K/N$. Both $y_\infty$ and $K$ are set by the
repeat-unit chemistry, through $\tilde{\mathbf{h}}^{\mathrm{poly}}$ and
the learned matrix $\mathbf{W}_{\mathrm{end}}$, which allows the magnitude
and sign of the correction to vary between polymers as $K$ does. The
crossover scale $N_0$ bounds the correction at low $N$, where chain ends
are no longer a dilute perturbation and the linear $1/M_n$ dependence of
$T_g$ breaks down; it is a learned scalar. For properties that follow
power laws in chain length, such as the zero-shear viscosity or the
radius of gyration, the natural coordinate is $\log N$, in which such
scalings are linear~\cite{rubinstein2003polymer}. The prediction head
therefore additionally receives $\log N$ as an input feature, and
crossovers between scaling regimes, such as the onset of entanglement,
are left to its nonlinear layers. Throughout, $N$ is the number-average
degree of polymerization, $N = M_n / M_0$, with $M_0$ the repeat-unit
mass. When the full molecular weight distribution is available,
\modelname{} instead incorporates it through a distribution-aware
encoding~\cite{hu2025mw} that captures dispersity and higher-order
moments.

% ---------------------------------------------------------------------------
\subsection*{Prediction head}
\label{sec:methods-prediction}
% ---------------------------------------------------------------------------

Polymer embedding
$\tilde{\mathbf{h}}^{\mathrm{poly}}_{\mathrm{MW}}$ is mapped to
property predictions through a residual
MLP~\cite{he2016deep}:
\begin{equation}
  \hat{\mathbf{y}} = \mathbf{W}_{\mathrm{out}} \, f_{\mathrm{ResNet}}(\mathbf{W}_{\mathrm{in}} \, \tilde{\mathbf{h}}^{\mathrm{poly}}_{\mathrm{MW}})
\end{equation}
where $f_{\mathrm{ResNet}}$ consists of residual blocks with PReLU
activations and
$\mathbf{W}_{\mathrm{out}} \in \mathbb{R}^{P \times d_h}$ maps to
the number of target properties $P$.

% ---------------------------------------------------------------------------
\subsection*{Variational regularization and generative pathway}
\label{sec:methods-vae}
% ---------------------------------------------------------------------------

The encoder shares its architecture with a variational autoencoder
(VAE)~\cite{kingma2014vae} pathway.
Root vectors are mapped to a latent distribution via learned mean and
variance networks:
\begin{equation}
  \boldsymbol{\mu} = \mathbf{W}_\mu \, \mathbf{h}^{\mathrm{root}}, \quad
  \log \boldsymbol{\sigma}^2 = -|\mathbf{W}_\sigma \, \mathbf{h}^{\mathrm{root}}|
\end{equation}
where the absolute-value constraint ensures $\sigma^2 \leq 1$.
Latent vectors $\mathbf{z}$ are sampled using the reparameterization
trick~\cite{kingma2014vae} and decoded by an autoregressive graph
decoder that reconstructs molecules node-by-node through the motif
vocabulary.
The KL divergence term
$D_{\mathrm{KL}}(q(\mathbf{z}|\mathbf{x}) \| p(\mathbf{z}))$ with
a unit Gaussian prior serves as an implicit regularizer for the
encoder, encouraging smooth, structured latent representations.
The learned latent space provides the foundation for the inverse design
 through latent space sampling and conditional generation, as demonstrated
in the PFAS case study discussed in the following (Fig.~\ref{fig:pfas}).

% ---------------------------------------------------------------------------
\subsection*{Loss function}
\label{sec:methods-loss}
% ---------------------------------------------------------------------------

The total training objective combines property prediction and
variational regularization:
\begin{equation}
  \mathcal{L} = \mathcal{L}_{\mathrm{prop}} + \lambda \cdot \beta_t \cdot D_{\mathrm{KL}}
\end{equation}
The property loss $\mathcal{L}_{\mathrm{prop}}$ is the sum of
the mean squared errors per-property with NaN masking:
\begin{equation}
  \mathcal{L}_{\mathrm{prop}} = \sum_{j=1}^{P} \frac{1}{|\mathcal{M}_j|} \sum_{i \in \mathcal{M}_j} (y_{ij} - \hat{y}_{ij})^2
\end{equation}
where $\mathcal{M}_j = \{i : y_{ij} \neq \text{NaN}\}$ is the set
of samples with valid labels for property $j$.
This enables multi-task learning with incomplete labels, which is
common in polymer datasets where different properties are measured on
different subsets of materials.

The KL weight follows a linear annealing schedule
$\beta_t = \beta_0 + t \cdot \Delta\beta$, clamped to
$[\beta_{\min}, \beta_{\max}]$, with an additional scaling factor
$\lambda$ (default $0.01$) to balance property prediction and
regularization.
% ---------------------------------------------------------------------------
\subsection*{Molecular Dynamics}
\label{sec:methods-generation}
% ---------------------------------------------------------------------------
\paragraph{Force Field Parameters and Equilibration Setup.}
Molecular Dynamics (MD) simulations were carried out using the OPLS force field. This force field was selected due to its well-documented accuracy in predicting the density and thermodynamic properties of organic liquids. Automatic force field parameterization of the monomers is performed using the same procedure as described in \cite{schneider2022}. An initial low density configuration is used and equilibrated in NVT at 600K. Then an NPT simulation in a temperature ramp from 600K to 300K in 50ns is used to reach the equilibrated density, followed by an additional NVT relaxation at 300K for 20ns. Simulations of these interfaces were performed using the \textbf{OpenMM} simulation engine.
\paragraph{Interfacial Tension Simulations}
To evaluate the interfacial behavior of the organic phase, biphasic systems were constructed by joining the equilibrated organic bulk (from the previous step) with pre-equilibrated boxes of \textbf{water, hexane, and air} (air represented as a vacuum gap).

The interfacial tension ($\gamma$) was calculated using the \textbf{Test Area 2 (TA2)} method. This non-exponential perturbative approach calculates the change in free energy ($\Delta A$) associated with a virtual change in the interfacial surface area ($\mathcal{A}$) at a constant volume. The interfacial tension is expressed as

\begin{equation}
\gamma = \lim_{\Delta \mathcal{A} \to 0} \left( \frac{\Delta A}{\Delta \mathcal{A}} \right)_{N,V,T} = \left< \frac{\Delta U}{\Delta \mathcal{A}} \right >_{N,V,T}
\end{equation}

where $\Delta U$ is the difference in potential energy between the perturbed and unperturbed states.

% ---------------------------------------------------------------------------
\subsection*{Training details}
\label{sec:methods-training}
% ---------------------------------------------------------------------------

\paragraph{Data splitting.}
All results are reported using five-fold cross-validation.
A composite stratification score is computed by normalizing each
property to $[0,1]$ and averaging, then discretized into five
quantile bins to ensure balanced property distributions across folds.
Within each fold, a held-out validation set (10\% of the training
partition) is reserved for early stopping and hyperparameter
selection.
All methods in Table~\ref{tab:baselines} are trained and evaluated
in identical splits.
The parity plots in Fig.~\ref{fig:architecture}c show predictions
from a representative fold.

\paragraph{Normalization.}
The target properties are normalized to the z-score using training-set
statistics:
$\tilde{y} = (y - \mu_{\mathrm{train}}) / \sigma_{\mathrm{train}}$.
Statistics are saved for reproducible inference.

\paragraph{Optimization.}
We use AdamW~\cite{loshchilov2017decoupled} with the learning rate $10^{-3}$,
the weight decay $10^{-4}$ (excluded from biases and normalization
layers) and the gradient clipping at max norm 5.0.
The learning rate follows a warmup-cosine schedule: linear warmup for
5 epochs, then cosine decay to $10^{-6}$.
Training runs for up to 100 epochs with early stopping based on
validation loss (patience of 55 epochs).

% ---------------------------------------------------------------------------
\subsection*{Ablation study design}
\label{sec:methods-ablation}
% ---------------------------------------------------------------------------

To quantify the contribution of each architectural component, we
performed a systematic ablation study in which exactly one component
is removed from the full model while keeping all the others intact.
The four ablated variants are:
\begin{enumerate}
  \item \textit{$-$ Composition-aware aggregation}: mole-fraction
    weighting (Eq.~\ref{eq:weighted-agg}) is replaced with simple
    mean pooling over fragment embeddings.
  \item \textit{$-$ Root vector (inject)}: the root vector pathway
    that modulates the prediction head is removed; only the
 embedding of the fragment aggregated by composition is used.
  \item \textit{$-$ Stochastic edge weighting}: \grins{}-derived bond
    probabilities at the monomer graph level are set to uniform
    ($\alpha_{v}^{(1)} = 1$ for all nodes).
  \item \textit{$-$ Root vector (concat)}: Concatenation of the root vector
 after removing the aggregation based on the composition.
\end{enumerate}
$T_g$ is used as the representative property for ablation because its
sensitivity to cooperative segmental dynamics amplifies the effect of
each design choice (see Results and Fig.~\ref{fig:architecture}d).

% ---------------------------------------------------------------------------
\subsection*{Evaluation metrics}
\label{sec:methods-metrics}
% ---------------------------------------------------------------------------

The performance of the model is assessed on the held-out test set using mean
absolute error (MAE), root mean squared error (RMSE), and the
coefficient of determination ($R^2$):
\begin{equation}
  R^2 = 1 - \frac{\sum_i (y_i - \hat{y}_i)^2}{\sum_i (y_i - \bar{y})^2}
\end{equation}
For the baseline comparison (Table~\ref{tab:baselines}), all methods
are evaluated under identical five-fold cross-validation splits, with
mean and standard deviation reported across folds.

\section*{Data availability}

The polymer datasets used for training were generated using our in-house molecular simulation framework. These datasets constitute the data supporting the findings of this study and are available upon request.

% ============================================================================
% CODE AVAILABILITY
% ============================================================================

\section*{Code availability}

The \modelname{} code and trained model weights are available upon request.

% ============================================================================
% ACKNOWLEDGEMENTS
% ============================================================================

%\section*{Acknowledgements}

% ============================================================================
% AUTHOR CONTRIBUTIONS
% ============================================================================

\section*{Author contributions}

J.J.d.P conceived and supervised the project.
G.S. developed the model and performed experiments.
G.Z and Y.T developed the \grins{} representation.
G.P.L, J.P., and D.S. performed molecular simulations.
All authors discussed the results and wrote the manuscript.

% ============================================================================
% COMPETING INTERESTS
% ============================================================================

\section*{Competing interests}
%\placeholder{ }
The authors declare no competing interests.

% ============================================================================
% REFERENCES
% ============================================================================

\printbibliography[title={References}]

% ============================================================================
% EXTENDED DATA
% ============================================================================

\clearpage
\appendix
\setcounter{figure}{0}
\setcounter{table}{0}
\renewcommand{\thefigure}{ED~\arabic{figure}}
\renewcommand{\thetable}{ED~Table~\arabic{table}}

\end{document}